\documentclass[]{acm_sen_article}

\usepackage{lipsum}
\usepackage{flushend}

\usepackage{graphicx} 
\PassOptionsToPackage{hyphens}{url}
\usepackage{hyperref}
\usepackage{xspace}
\usepackage[numbers]{natbib}
\usepackage{balance}
\usepackage{xcolor}
\hypersetup{
  colorlinks=false,
  pdfborder={0 0 0}
}

\usepackage{fontawesome5}
\usepackage{framed}
\definecolor{formalshadelight}{RGB}{242,242,242}
\definecolor{formalshadedark}{RGB}{166,166,166}

\usepackage{csquotes}

\begin{document}

\title{Empirical Software Engineering in Practice: \\Insights from Google}

\subtitle{ACM SIGSOFT SEN-ESE Column}

\numberofauthors{2} 
\author{
Roberto Verdecchia\\
       \affaddr{University of Florence}\\
       \affaddr{Italy}\\
       \email{roberto.verdecchia@unifi.it}
\and
Justus Bogner\\
       \affaddr{Vrije Universiteit Amsterdam}\\
       \affaddr{The Netherlands}\\
       \email{j.bogner@vu.nl}
}

\maketitle

\begin{abstract}
While it is fairly well known how empirical software engineering (ESE) is used in the academic world, we have limited knowledge of how ESE is practiced in industry.
As part of our regular column on empirical software engineering (ACM SIGSOFT SEN-ESE), we want to dedicate a series of articles to interviewing ESE practitioners from various companies.
Among other things, we want to understand how ESE processes are implemented in industry, e.g., different research methods, how practitioners decide on what to study, how research results are used within companies and beyond, and if they face recurrent impediments to using ESE methods in industrial contexts.
In the first edition of \textit{ESE in Practice}, we are joined by Ciera Jaspan and Collin Green from the Developer Intelligence team at Google.
This article is a faithful account of our conversation from August 13, 2026, which we edited for our~column.
\end{abstract}

\section{ESE at Google}

\bigskip

\textbf{Moderators: To start things off, we kindly ask you both to briefly introduce yourselves. Who are you, what's your background, and what do you do in your current role?}

\textbf{CJ:} I'm Ciera Jaspan\footnote{\url{https://research.google/people/cierajaspan}}, a software engineer and researcher within our team here at Google, where we study engineering productivity. Before this, I primarily worked on static analysis tools, software architecture, and software frameworks, but that was about 10 years ago. I've always been in the software development tooling space throughout my PhD and even before that to some degree.

\textbf{CG:} I'm Collin Green.\footnote{\url{https://research.google/people/107023}} I am not an engineer by any means. I'm a psychologist by training, meaning I started in academic cognitive psychology and then slowly moved through different technical domains in my work. Today, I do user experience (UX) research, and at Google, we treat our engineers as users of our products and users of our systems. My first UX research job was actually at NASA, where we studied how software could be better designed for engineers in the space program. After that, I worked on similar things but for radiation oncologists, medical physicists, and radiotherapists, helping them do their jobs. And for the last nine years, I've been learning about and studying software engineers at Google. I try to help them be happy and productive by providing a behavioral science and social science perspective to the research that we do.

\textbf{Moderators: Thank you both! Can you please describe what your team does in a few more details?}

\textbf{CJ:} We try to understand what makes software engineers happy and ultimately productive. Our team is not a team that builds developer tools. We have a huge number of teams at Google that do that. They're responsible for our software frameworks, plus the processes and infrastructure that we use. But our team is more of a research team spanning across those. We try to understand what the largest pain points for software engineers are right now, i.e., what is the biggest opportunity that we have to improve their developer experience and to make them more efficient? We then provide these results as feedback to the tools teams, and we help them with experiments. For example, if they've got a new intervention, we help them understand what the impact is and if it’s worth continuing. Such an intervention could be anything from an added tool feature to a whole new tool. We even worked with our onboarding team to set up an AB experiment for new people joining Google to study how long engineers take to ramp up. We provided them with the metrics and data to be able to understand whether their new intervention was worth it or not.

\textbf{CG:} What Ciera just said is really key because our team's mission has somewhat evolved in the last decade. But providing metrics to help assess interventions and process or tooling changes has always been a key piece of what we do. Related to this is deciding whether we build a metric that's fast, convenient, and easy to understand or one that's slow and harder to understand but also more generalizable and rigorous. This spectrum is something that we're always dancing across for the metrics that Ciera just mentioned. We spent a bunch of time trying to understand what \enquote{ramping up} actually means in terms of the tasks that engineers do and the things that they need to learn. How should we even measure when somebody is ramped up according to their own subjective experience? And then we tried to figure out suitable proxies for that ramp-up that we can observe at scale using objective data. Finding scalable, objective metrics that tell us something that's hard to observe in the world is, I think, really a key challenge for us. We need to balance the convenience and speed of counting things that are easy to count with counting actually meaningful things that reflect our engineers’ experiences and the actual operation of the huge sociotechnical system that is Google engineering. That's the core intention of our team.

\textbf{Moderators: Awesome, thank you! Our column is all about Empirical Software Engineering (ESE), so we are very interested in what role empiricism plays in your job, e.g., what research methods you use in your work.}

\textbf{CJ:} So, we've got a lot of method geeks and have explicitly hired people over time who expand our method toolbox. When I joined this team, I had some research methods in my pocket from my PhD work, but by working with people like Collin and other UX researchers from other fields, I really started learning about other types of methods. For example, one of our first UX researchers introduced me to ethnographic research. She was a true specialist and got really into it. And I thought, okay, that's a new method that we now can use when we need to. We always try to apply research methods based on the problem we have. When we get a new problem, we start by asking what the research question is and what our hypotheses are. We also think about who needs this answer and what kind of methods they consider valid. I think that may be a difference between industrial and academic research. We need to help some concrete person make a decision. If this person doesn't believe, e.g., in survey data or certain types of log data, we have to take that into account. We then adjust our method accordingly to provide them with what they are going to trust so that they actually can take action on it. We're not saying that we're giving them the answer they want. However, we're trying to pick a method that we know they're going to believe in so that they don't debate it later. This is one of the reasons we need a wide variety of research methods, but we also receive a wide variety of questions. Lastly, there are some limitations for what kind of research we can do, e.g., in terms of required speed vs. extremely rigorous, slow research.

\textbf{CG:} Yes. The other thing that's really key here, which is another reason why we added people with methodological breadth to the team, is that we can't just run arbitrary experiments, right? Google engineers are doing many important things, and we can't just say 10\% of engineers will do it this way for a study. I mean, to be clear, there have been examples of where we have run experiments like this. At one point, we were assessing whether adding higher-performance build machines would change the way engineers behaved and how their productivity would look. We actually ran a double-blind experiment on this, but this is definitely rare. A lot of what we do is essentially observational research, or we have to work with data that's already being collected. We somehow have to work with not being able to run fully controlled experiments most of the time. It has actually more in common with things like economics, public health, or clinical research, where you don't have the entire space of combinations of conditions that you can implement. You have to work with the controls and conditions you have. We therefore do a fair bit of pre-post analysis, or we do propensity matching to find engineers similar to the people who did something differently to see if their outcomes also are different. So, those methods from economics, sociology, or more broadly speaking, from data science all come into play because we can't always run proper experiments. We have to work in a context where Google engineers are doing their jobs, and we just need to understand how they're changing. We've had economists on the team and a bunch of psychologists. We also have someone with a political science background or from engineering or public policy. The variety of PhD disciplines in our team is quite large.

\textbf{Moderators: So, you really use all kinds of methods, from ethnography, survey research, and repository mining, to experiments?}

\textbf{CG:} Yes. Like Ciera said, we're pretty pragmatic about what methods we apply. I think one of the reasons we have a lot of people from the social and behavioral sciences is that Google is like a big sociotechnical system. That means it's somewhat amenable to the same research methods that are used to study things like economics. Psychologists are also used to trying to measure things that are fundamentally unobservable. They try to infer things about these unobservable states from data. I think social and behavioral scientists are also good at understanding that they're only measuring a tiny little slice of what's actually happening. So, they’re used to keeping their conclusions, assertions, and methods consistent with the fact that we don't know everything about what's happening. We see just a little fraction, a little reflection.

\textbf{Moderators: Fascinating. This nicely feeds into our next question. You seem to nearly always combine quantitative and qualitative methods. Can you elaborate a bit on why you do this and why you think it’s important?}

\textbf{CJ:} It's critical to what we do. We often say that the quantitative part tells you the \enquote{what}, but it does not tell you the \enquote{why}. The qualitative part is what tells you the \enquote{why}. Let me give you an example. The median time it takes people at Google to review a code change is five minutes. Cool. I can give you that number, but that tells you nothing about whether it is good or bad. It's just a number. It doesn't tell us whether it should be slower or faster. Are we doing good here? Is this a pain point for software engineers? Those are answers that you get by talking to the people to find out whether this is a concern or not. For example, we've seen that number change during COVID. Okay, again, good or bad? Was this a problem? All we knew was that something happened, and now we have to go and explore why. That's why we need to have both types of methods. When we do mixed-method research, it also helps us understand whether our qualitative and quantitative methods are giving us the same picture. We try to triangulate all the time. And when everything tells us the same thing—our survey data, our interview data, and our log data—it gives us higher confidence that this is in fact a real phenomenon. By the way, this also helps with those stakeholders who might not believe one of those three things. But we do have times when they do not agree. Then that just means that we need to dig in deeper. For example, when survey interviews and our log data don't agree, it's frequently because of errors in our log data that we were not aware of. We've had that happen several times. We've also had cases where our survey was actually answering a different question than the log data. They were simply looking at two different things, and we thought they were looking at the same thing. But when you talk to people via interviews, you find out that they have a different idea in their head when answering the survey question than what we thought. That’s useful to know. It's not that there's disagreement because the methods are wrong. It's a disagreement because we unknowingly measured two different~things.

\textbf{CG:} Frequently, this happens because the logs are too narrow. For example, if we're trying to understand what makes a team productive, we can look at processes, tools, or the performance of infrastructure as predictors for that. Sometimes, we may get an interesting signal, but a lot of the variance won't be accounted for by the thing that we happen to be measuring with logs. Everything from team dynamics, which are still specific to the work environment, to weird externalities will matter. For example, when the World Cup happened, that actually had a measurable impact on productivity at certain moments in certain locations for Google engineering. The pandemic is another great example of externalities that influence productivity actually as much or more than engineering interventions. And if you don't do qualitative work or you don't do survey work, you're ignoring the major drivers of what's happening. During the presidential elections, we've seen measurable effects on engineering productivity because people are busy talking and thinking about what's happening in their world. That's impacting their productivity more than which tool they're using. So, treating qualitative, behavioral, and survey research as a way to scoop up a bunch of stuff for which we don't have good, objective measurements counts for a lot in the performance of the organization. Additionally, there's a bunch of stuff that we simply don't know how to count. An example that I've used a lot is that we really care whether engineers are in flow or not, whether they're focused and highly productive in a certain moment. That's been a slippery concept to nail down from a psychological perspective, but definitely from an objective perspective. We used qualitative and behavioral methods to get a handle on what flow looks like subjectively and qualitatively. Then we used that as a benchmark for trying to measure this same construct objectively. Usually, we want to measure things with log data because it scales, and it's passive and doesn't bother the engineers. We can do it continuously for everyone and get a good longitudinal picture in a big data set. But to get there, we need to know that we're measuring something that actually is meaningful and exists in the world. For flow, we did a bunch of research on what it looks like qualitatively~\cite{Brown2023}. What are the things that drive it up and down? And then we essentially started to build a mathematical model of how to detect it. We used our subjective survey data to see if our model is any good. Sometimes, it's that back and forth that actually helps you bootstrap from not knowing how to measure or count this to finding a suitable proxy for the thing that we actually want to measure. We also did this successfully with ramp-up~\cite{Green2023}. We tried to do that with technical debt but didn't succeed in that case.

\textbf{CJ:} We didn't succeed with code quality either. We have multiple golden data sets that we curated from our surveys, interviews, and various other diary studies, but we still don't have a good logs-based metric for that one.

\textbf{Moderators: Very interesting to hear about the importance of qualitative methods in your work! Moving on: You hinted a bit about it in previous answers, but maybe you can describe a bit more where your research questions usually come from. Is it from management, engineers, or your own team?}

\textbf{CJ:} It's a mix. We also have a dedicated, quarterly survey on this, which gives us some information about the problems developers are facing. We have used that before to identify a top problem that engineers are now reporting to us through our survey, and then we're going to go look into that. We also get questions from Google-wide leadership about topics that they are concerned about. Sometimes, these are things happening in the world, e.g., COVID was a big one, or things that they are wrestling with today. For example, they're trying to make a decision around how to structure their organization or what types of efforts they should put people on. So, we're getting our research questions from multiple sources. We also have more than we have time to answer, meaning we frequently end up having to prioritize. That prioritization process is a little bit messy, but it's ultimately about how much impact we think answering a question will have and if we are uniquely positioned to be able to answer it. If there's an answer that's already out there, we can refer to an external paper. We also certainly would prioritize a senior VP at Google facing a big decision that would affect thousands of engineers at the company vs. someone who's just an independent contributor contemplating something they think would be cool to understand. The other big factor we use is whether there is, in fact, a real decision to be made. We really try hard not to do any research just to prove someone's point about something they were going to do anyway. So, one of our favorite questions to ask our stakeholders is what they think is going to happen if our research results don’t align with their plans. If we get a negative result and find out that their hypothesis is completely wrong or if the results are inconclusive, how will that change their plans? Because if it's not going to change their plans, then there's no point in doing this research.

\textbf{CG:} I just want to add that the way we select research questions has definitely oscillated over time. Sometimes, we lean much more into leaders asking questions because they're trying to make these big decisions for the company, and we want to help inform those. COVID times were rife with these questions, e.g., how do distributed teams or hybrid work impact productivity? But then there are also times when we're much more getting research questions by trying to amplify the voices of our engineers. That’s the survey Ciera was talking about, where we hear about the biggest thing slowing people down. Sometimes, we try to put a definition around that challenge to make it visible to leadership and to just help people feel heard and to maybe plant the seeds that this is a problem to be solved. Because, actually, Googlers are great at taking the initiative and trying to solve a problem that they see. There's a culture of trying to have impact at the company. So, I think it's those two things: how can we amplify what we're hearing from the engineering population, and how can we be responsive to leaders who are asking for answers to important questions? That's been a bit of a dance over time, and I think even across projects, we will simultaneously have one project that's about amplifying what engineers are feeling, hearing, and experiencing and one trying to answer a leadership question.

\textbf{CJ:} And sometimes, it's both at once. I mean, technical debt is the one that's been alternating between the two. Either leadership discovers that tech debt is a problem or engineers are saying that tech debt is currently terrible, sometimes both at once. That's why that one keeps coming back as a problem.

\textbf{Moderators: Thank you! A question about your research results: When you finish a study, how are your results usually used within the company, and how do you also decide what to publish externally?}

\textbf{CJ:} We try to have that answer before we even do the research, if possible. Ideally, the results will be used by this person to make a particular decision. For external publications, the way I like to think about it is that we try not to chase publications because we are not in an academic setting. However, about 6 weeks before the ICSE deadline\footnote{\url{https://conf.researchr.org/series/icse}}, we take some time and look back at what we've done in the past year to see if there is anything of interest. We have a pretty high bar for what we publish. We do not publish at any lower-tier conferences or workshops. It's just not worth our time. It is a little bit like \enquote{go for the top or don't bother}. We also ask if this is interesting to someone outside of Google. Can it have a high enough impact to be at a top-tier conference? And then, of course, are we actually allowed to publish it? Most of the time in our domain, these things are publishable. It is actually within Google's best interest to tell people about what we have learned about code quality, technical debt, or ramp-up. We want that message out there because it helps the rest of the industry as well, and we want other academics working on this, too. Sometimes, we just want academics to know that this is a problem we are facing. We worked on it a bit, and now someone else can take it away and continue so that we can work on something else.

\textbf{CG:} I think, even though Ciera is correct that we ideally want to know how a result is going to be used before we do the work (that is the optimal situation), we don't always have a super concrete idea. Especially when we've unearthed a foundational question that we think is interesting but nobody's actually asking us yet. Sometimes, we will chase those questions when we're thinking about a problem, and there'll be a sort of underlying foundational problem. Can we solve that foundational problem? That does happen irrespective of whether we have a specific target for a piece of research. Additionally, we often try to cast a fairly wide net in terms of disseminating the research inside the company. We send out our research in reports, presentations, or dashboards. Sometimes, we even publish data tables and scripts. We send it to the many people who have signed up to hear about our research results. In some cases, a person we didn't even know existed will reach out and pick it up, e.g., they are interested in a specific facet or like to apply it. That does happen. As another motivation for this internal publishing, we try to be really transparent with our research. We don't want to be seen as some surveillance team at Google, even though we have a lot of data about our employees. We want engineers and tech workers in general to view us as working in their interest and working to amplify their voices and their experiences. Being really transparent about what we do is very important for this. There have been a few cases where we had to delay or narrow down what we were sharing, particularly on sensitive topics. But in general, we try to be quite transparent with what we've done and share it widely within the company. As Ciera said, we're somewhat opportunistic about what we publish externally. Google does have a very large population of engineers and a lot of granular data about them, more than many other companies. People are usually interested in what we have to say, but we never select a research question because it will lead to a publication. It's always a \enquote{nice to have} that we sometimes take advantage of, but it's not a driver for what we select.

\textbf{CJ:} Collin mentioned something there about research projects where we think that there's going to be people interested in the results; we just don't know who they are yet. Many times, those projects start because we have seen something happen in the past. The work we did on flow was one of these. Four teams had come to us independently and wanted to measure the impact of a new thing they were building. We asked what their ultimate goal was, and they all said they're trying to improve flow. Of course, they wanted to know if we have a metric for flow. We said we could build one, but it's going to take us at least nine months. Usually, they want to launch like next week and are not willing to wait. But once this happened about four times, we decided to build it ourselves because there was clearly a need. So, in such cases, we just suck up the cost and pay for it ourselves for future teams.

\textbf{CG:} Yeah, when we can. Code quality is the great counterexample. We would love to have a solid code quality metric. Everybody wants one. They all want to be able to show what they bought in speed without adversely impacting quality. But that problem—how to objectively measure code quality—has been elusive. We certainly would publish such a result externally if we had a good one. So far, we only got negative results.

\textbf{Moderators: Very interesting. Moving on, when applying your research results, do you sometimes face challenges or impediments, e.g., skepticism, a lack of participation, or a lack of tangible impact?}

\textbf{CG:} Yeah, I mean, skepticism for sure. I think that's the nature of any research. Maybe it's a little less present in an academic environment, but in industry, there are many people with interests in outcomes. If you show up and tell them that your data suggests that their desired outcome is not true, there’s obviously some resistance. I mean, people here want to make good decisions and are generally rational. But that doesn't mean they don't have any biases and beliefs. So, of course, we face more scrutiny when we have an unpopular conclusion, just like anyone would. We already mentioned that some people are more skeptical of certain methods than others. We once had a leader telling us basically that our survey data was just people's opinions. \textit{(laughs)} I mean, yeah, that's not wrong. But also, it's systematically gathered opinion at a very large scale. You listen to people's opinions all the time. Why aren't you listening to these? So, we face some methodological skepticism that comes as a complement to that. I don't know if we really see resistance. If there's a pretty unambiguously good decision or a clear outcome, I don't think we face resistance to implementing it. Sometimes, it's not feasible. That's an issue. And sometimes, it's hard to set a priority around something that would have a positive impact, but it's very long-term value, or it's a very small slice of the population that would be positively impacted.

\textbf{CJ:} We've definitely had times when people didn't believe our data for various reasons because it didn't match their preconceived notions. But again, this is also where using multiple methods helps. For example, we had a case where we presented some log data to a person, and she just straight up said this can't be right. Our data would be wrong because it didn't match the preconceived notions of her experience. But her experience was, as it turned out, not the normal experience and a little bit out of date. So, Collin and I went back to some complementary, qualitative diary data, which was super helpful here. We had the log data that said this thing usually takes three minutes. She didn't believe this and was convinced it would take at least 20 minutes. So, we said, let's look at the diaries. We pulled up examples where people were saying what they were doing, and then we looked at their logs. We just gave her one example after another. This one took two minutes for the engineer, and this is what they were doing. This one took 30 seconds, and this is what the engineer was doing. After three of those examples, she came around. She just needed to see that her experience was not the typical one.

\textbf{CG:} This is a great example. The match of the conclusion to people's intuition or presumption is really the driver for skepticism and questioning results. In many cases, it's not the method or nature of the data. If you tell people that you did a small-sample ethnographic study and it turned up exactly what they thought or hoped for, they're on board with ethnographic methods, or vice versa. \textit{(laughs)} I think the person Ciera mentioned was a very quantitative person, but the log data simply flew in the face of her expectations. The qualitative data were what brought her around. In most cases, I would have guessed the opposite was going to happen, namely that she wasn’t interested in us \enquote{telling stories} with qualitative data but trusted the numbers. But it was very much about expected vs. actual results here.

\textbf{Moderators: Nice example of the power of mixed-methods research. Another thing that we are wondering about: How are ESE or empirical methods actually used within non-research units at Google? For example, do product teams also use research methods or evidence-based practices like continuous experimentation?}

\textbf{CJ:} Yeah. All the time.

\textbf{CG:} I think there's two big ways. I let Ciera speak to the more formal product experimentation part of it because I think she can better describe it. As a UX researcher, I do very researchy and foundational things at Google. But there are many folks like me whose whole goal is to understand user needs and usability. They gather evidence about how users are interacting with our products, how the products are working for them, and what the products are doing well, which they provide to product teams to drive better design. That's how UX and HCI have been evolving for decades now. I think that is very common, this sort of data-driven design and empirical user research. I'm using \enquote{empirical} a little bit loosely here, but systematic user research is a part of almost every product team at Google. For the continuous experimentation part, I think Ciera can explain this better.

\textbf{CJ:} I mean, our team is focused on just the internal developer tools, but the methods we are using are used across Google. They are used in search, they are used in our cloud products, they're used on Android. Teams constantly try out something new, do an AB experiment by rolling this out to 5\% of the users, and then see what happens. Every single team that owns a product at Google has their suite of metrics that they track. They usually also have some qualitative data they try to grab, but that's way more expensive. I think we have an easier time getting access to that type of data internally because our users—the engineers—literally sit next to us. I definitely simply biked over and talked to somebody before. That’s a little harder when you have customers who are outside the company. But they still do that. They are constantly trying to gather data and look at it in different ways.

\textbf{CG:} Yeah, I think Google famously got very experiment-oriented early on. There's this true story about testing which of 41 shades of blue was optimal in a specific UI~\cite{Kohavi2013}. I think you can also go overboard with this. Today, some of the challenges that exist are more about coordinating experiments, designing them efficiently, and preventing collisions. We run so many experiments that actually making sure you're not blowing up one experiment by inserting another that overlaps is a big part of what we do. We also need to think about how we optimize those experiments. Because we have billions of users on some products, we obviously don't want to run an experiment with half the population. But what is the sweet spot where you get enough data fast enough to make a good decision but also don't suffer negative consequences from having experimented with something that actually made performance or experience worse? I think that’s where a lot of the hard problems are at this point. It's still more in the design than engineering domain, though you can also do performance experiments and similar things.

\textbf{CJ:} There are some fun engineering problems, too. This is not my domain, but some teams have problems where even rolling out the experiment affects the non-experiment groups because there's some back-training that happens. Plus, you can have cases where your experiment looked good when it was rolled out to 1\% of users, but it looks terrible when it was rolled out to 5\% of users. Often, this happens because the control group changed because it was learning from the first experiment. It's an engineering challenge to figure out how to keep your groups separate and how to adjust for the fact that running the experiment could actually change the results in both groups.

\textbf{CG:} There's also this \enquote{always-on problem}: let’s say you ran an experiment for four weeks, and it looked good. But there's a novelty effect, and four weeks after that, it looks terrible. For this, you can do things like a holdback sample. You run your experiment with 1\%. It looks good, so you broadly deploy it, but you still hold back 1\% in a control condition so that you can see whether you have a deleterious effect over time. Because of these complexities, we have so many data scientists and data engineers. These are definitely bread-and-butter practices for most product development teams at Google.

\textbf{Moderators: Very nice, it means that empirical methods are wide\-spread throughout the whole organization. Another question is about how you perceive academic research. Is most academic research actually useful for you, or is your industry context so different that you don't get a lot of value from academic publications?}

\textbf{CJ:} Okay, that's complicated. \textit{(laughs)} We do actually get a lot from academic publications. I mean, there's a reason why I go to ICSE every year. I always come back with at least some new research ideas or things to try out, but also things that I know now aren't going to work, so I'm not going to bother with them. I always come back with something interesting. However, there is definitely opportunity for improvement. I spend a lot of time talking to students, where I tell them to think bigger and bolder or to stop salami-slicing their research. And I want to see the negative results, too. I think there's a lot of emphasis on trying to reach positive results. Then you see students submitting these papers that are just barely positive or on some niche problem that I don't really care about, but it happened to work for that area. I just did a literature review a couple of weeks ago. I was experimenting with a new potential code quality metric and wanted to figure out what people had done already. I found some work that looked promising, but it did not replicate internally. That wasn't the fault of those researchers. It was the data set that they had access to and for which it happened to work. It gave me an idea, and I liked what they had done. I could take this core idea and try to make it work on a larger, more realistic industrial dataset, where it sadly didn’t work. And that’s okay.

\textbf{CG:} I won't speak for Ciera, but I think I'm an academic at heart, even though I left academia a long time ago. I love research to the extent that it's probably career-limiting in some capacity because I would rather get it right and truly understand something than say the palatable thing or act politically savvy. \textit{(laughs)} I love to hear idea generation from academic studies, like what sort of models they're proposing. A thing that has been a staple for me in my career is trying to bring things from other domains and applying them to the problems that we're trying to solve. I've done that a lot with psychology. For example, the way psychologists discuss interrater reliability shaped how we evaluate some of the data used for understanding and validating our metrics. So, both methodological analogy and just thinking about how models that work in another field apply over here are important to me. A current example is that I'm rethinking some of our ramp-up work by looking at models they apply in medicine to understand the efficacy of a treatment plan. They're interested in how long it takes to get to some critical event—usually a negative one like the patient dying, or a positive one like the tumor going away. But they have the same complexities in those data sets that we sometimes face in understanding how fast people learn. So, taking models that work in other domains and bringing them to our work is a big thing that I seek to do actively because I think it's overlooked in general. Looking at the academic literature and looking at how people are applying some of the concepts, models, or analytical methods in other fields is really enriching for me. I don't read as much as I would like, but often, I find a key paper or an idea that I think translates really nicely. That gives me a new angle on a problem that we have. And there's a lot of value in that, in my opinion.

\textbf{CJ:} Collin's ability to find interesting papers in other fields is impressive. He often ends up pulling something interesting from philosophy, economics, or psychology where I wonder how he even found that. But it ends up being exactly the thing that we needed.

\textbf{CG:} Yeah, my signal-to-noise ratio is maybe a little iffy, but I do have some hits. \textit{(laughs)}

\textbf{Moderators: Sounds very nice! To round things off, two more questions looking towards the future. First one: Where do you think the future of ESE is heading, or where do you wish it would go?}

\textbf{CJ:} Hmm, I am not sure where it will go. Where I wish it will go is a little easier. I have seen a couple of problems with our field when it comes to using different methods. One of them is that we definitely have a sense within software engineering that there's always a right way of doing something. I think we have all seen this with reviews we get for our papers. People will complain that we didn't do it exactly this way, the only true right way of doing this method. One of the things that I've liked about being on my current team is finding out that that's not true at all. There is not one right way of doing every single method. Early on, we had a paper that I was struggling with internally. We used a research method, and the other software engineers were saying this isn't the right way of doing it. But we had just hired a teammate from behavioral economics. She looked at it and said that there are three ways of doing that, and in this case, they're all valid. She just had a very different perspective on this. I think we're very rigid in software engineering. We tend to think that there is a right way of using every single method, as opposed to looking more holistically at what a problem requires and what makes sense for this issue. I'd like to see us take a more holistic approach the way the social sciences do. I'd also like to see us be able to publish our mixed-methods results more easily. One of the difficult things we've run into a couple of times is that conference paper page limits are unsuitable for mixed-methods results. If you have to describe in one paper an interview study, a survey, and a quantitative analysis, you just do not have the pages to do that. We've tried doing that, but then you have to omit many details our field expects, again, because we're so rigid about things. If you have a survey, you must report certain things, and then you run out of pages. So, I would like to see us find a way of being able to publish that type of work better. I don't want to salami-slice such research because it's not interesting if I split it into three. Each of those three parts on its own is actually not interesting; it is the combination of them that is the interesting piece.\footnote{See also the related Call to Action by \citet{Storey2025}.}

\textbf{CG:} Ciera and I gave a talk and have written an article~\cite{Jaspan2025} about this famous quote: \enquote{All models are wrong, but some are useful}. I think that gets right to what Ciera's talking about. Whatever technique or research method we choose, it has merits, but it is almost guaranteed to be incomplete and/or wrong in some way.\footnote{This mirrors nicely the underlying philosophy of the ABC framework by \citet{Stol2018}.} We should be embracing that and accepting that we are always looking at slices or facets of the problem, especially for something like software engineering in practice, which is like a big messy human system. There's just not going to be a single answer in most cases. People joke that, whenever they ask a UX researcher a question, the answer is always \enquote{it depends}. And it does; it always depends on the context, the use case, the situation, etc. So, I would like to see more people acknowledge, as Ciera said, that there's not a single right way. Whatever model you adopt, it may be useful for your context, but it is not the be-all and end-all definitive answer to life and the universe and everything. Another thing I would like to see is that we should try more to incorporate the non-technical factors into the way software engineering works. Understanding team dynamics, communication, psychological safety, or prioritization processes has a massive impact on the effectiveness of a software engineering organization. They don't have anything to do with the technical correctness of the code. Instead, they are about knowing what you're building, why, who it's for, what it's supposed to do, and what success looks like. I know those aren't software engineering problems per se, but they are wrapped around software engineering as a discipline. It is inextricable from those things, and I would like to see the empirical lens expand to encompass most of those things.

\textbf{Moderators: Great! We’ve reached the final question: in your opinion, what are some of the biggest open questions in empirical software engineering?}

\textbf{CG:} I don't know if I can answer this question. I'm gonna play the \enquote{I’m not an engineer} card. \textit{(laughs)}

\textbf{CJ:} Okay. I mean, every time I give an academic talk, I have a slide at the end with my top things that someone needs to answer. So, I'll pull on that. We still have no useful code quality metrics at all. They're all terrible. They all don't reflect what developers think of as quality. We also still have no way of measuring the velocity of software development across the entire life cycle. We have velocity metrics for small chunks of it. For example, I can measure the time it takes for a code review to go through our process, or I can measure the time from when a piece of code is submitted to when it's launched in production. But there's a need to have an overall measure, from the first time someone has an idea to build something to the time it is rolled out to 100\% of users or ends up dying at some point. How many of those ideas die, when do they die, and how far do they get? How much time have we sunk into them? We need this overall measure of velocity. I think we also need to better understand how people collaborate, especially with AI coming online now. There's a lot of really interesting stuff happening with how AI is changing collaboration. Collaboration was already something we needed to understand better before. How do developers collaborate? How do we organize highly functioning teams? AI has definitely changed things in interesting and new ways because how do we collaborate with the robot now? How does the bot collaborate with other bots? And how does that change how we collaborate with each other? One of our interns was just telling us about how she is seeing that the existence of AI changes how junior and senior engineers collaborate. The junior engineers don't ask the senior engineers all the little \enquote{how} questions anymore; they ask those to the AI. But she says that's getting replaced by junior engineers asking more \enquote{why} questions to the senior engineers, which is actually great. We want that. We want them to be having those discussions of why we are building this in the first place. What's the strategy here? What's the vision? So, that's pretty cool. However, I'd like to know more about what's going on there and how that changes how we make teams and support high-functioning groups of~people.

\textbf{Moderators: Very interesting. Collin, do you want to add something, or do you keep pulling your \enquote{I'm not an engineer} card?}

\textbf{CG:} Ciera's already ventured into the not-so-much engineering territory a little bit with her last point. I definitely agree. How do we optimize software engineering teams? What is the role for humans going to look like in the future? That's everything from what is an efficient thing to delegate to automation vs. to reserve for humans to how do we set human beings up to be maximally productive, efficient, and creative at the things that are essentially human? Whether it's creativity, prioritization, or judging quality, I don't know exactly what the niche for human beings will be in the future. It's probably some combination of those things. But how do we structure organizations, teams, and set individuals up for success if that is the niche that they're filling? Many folks in tech are acting like AI and automation are brand new and they emerged two years ago and have never been thought of before. And a lot of it is brand new for sure. But the automation literature is extensive. It goes back decades. People have been doing things like setting up autopilot in commercial airliners, trying to build self-driving cars, or automating industrial processes in factories for a very long time. So, I think we should try to learn from that. What do we automate? How? What is the humans’ new role? How does supervision override automation? How can we grant autonomy to automation but also reserve the right role for humans so that they can be accountable, take pride in their work, and do their work with maximum quality? Those are interesting questions, and I may have reserved them before because they may not seem like empirical software engineering questions. But in many ways, they are certainly relevant for ESE. We must understand the human role when software engineering is increasingly automated because that’s what’s going to happen. How do we set engineers up to deliver value in that new~environment?

\textbf{CJ:} That even comes back to education. I was just having a conversation yesterday where someone was asking about how to teach software engineers now that AI is doing everything. We have a set of critical tasks that we know our software engineers do. Currently, AI is doing 3 of these 30 critical tasks. And unfortunately, university education has historically been very focused on teaching students exactly those three tasks. But there's a whole bunch of stuff that no one's teaching students that we've been doing the whole time. Some of those will definitely be automated, but some won’t. We need to figure out what those are and how to better educate new people. So, there's going to be a whole new stream of empirical software engineering to figure out how to teach engineers because they need to be engineers now, not~programmers.

\textbf{Moderators: Thank you both so much for your time and the interesting discussion!}

\section{Conclusion}
We found it difficult to shorten our conversation with Ciera and Collin because their insights were just so interesting and relevant for the community.
Even though the research environment and context at Google are very particular, many of the takeaways from this interview should resonate with the ESE community.
Ciera and Collin highlighted the benefit of having a \textbf{plethora of different empirical research methods} at your disposal, which allows them to adjust to the different problems they face.
They also emphasized the importance of \textbf{mixed-method research and triangulation}, especially the combination of quantitative (the \enquote{what}) and qualitative data (the \enquote{why}).
Using multiple sources not only gives you more confidence in the results if they all align but can also guard against stakeholders or reviewers with method-related skepticism or preconceived notions of what the results should look like.
Several aspects of their way of publishing research also seem like solid advice for the general ESE community: \textbf{publish negative results}; aim for \textbf{bold and challenging problems} (not highly specialized niche cases); \textbf{publish holistic full papers} (stop salami-slicing your research).
However, we also need to remember that early-career researchers, especially PhD students, have other factors and constraints to work around, which may make it more risky for them to follow this advice to the letter.
Moreover, research stakeholders being biased towards certain empirical methods seems equivalent to method-related bias of ESE peer reviewers~\cite{Bogner2026}, which points to the need for \textbf{carefully selecting reviewers based on the research methods they are familiar with}.
Lastly, our community definitely should \textbf{reflect on our level of methodological rigidity and dogma} compared to other research disciplines.
Yes, methodological rigor is important, but we should not lose sight of the big picture and that there is usually more than one acceptable way of doing things.
Even the checklist-based approach of the Empirical Standards for Software Engineering Research~\cite{Ralph2020} clearly acknowledges that there are  acceptable deviations: \enquote{\textit{Empirical standards do not replace expert judgment with inflexible rubrics.}}

\section*{Acknowledgements}
We kindly thank Ciera Jaspan and Collin Green for their valuable time and insights!

\bibliographystyle{plainnat}
\bibliography{biblio}

\end{document}